\documentclass[prl,aps,amssymb,showpacs,twocolumn]{revtex4}

\usepackage{dcolumn}
\usepackage{graphicx}
\usepackage{bm}

\usepackage{amsmath}
\usepackage{amssymb}
\usepackage{amsthm}
\usepackage{amsfonts}
\usepackage{enumerate}
\usepackage{latexsym}

\input{epsf}

\begin{document}
\title{Topological Mott Insulators}
\author{S. Raghu$^{1}$, Xiao-Liang Qi$^1$, C. Honerkamp$^2$, and Shou-Cheng Zhang$^1$}
\affiliation{$^1$Department of Physics, McCullough Building,
Stanford University, Stanford, CA 94305-4045}
\affiliation{$^2$Theoretical Physics, Universit\"{a}t W\"{u}rzburg, D-97074 W\"{u}rzburg, Germany}
\date{\today}
\begin{abstract}
We consider extended Hubbard models with repulsive interactions on
a Honeycomb lattice and the transitions from the semi-metal phase
at half-filling to Mott insulating phases.  In particular, due to
the frustrating nature of the second-neighbor repulsive
interactions, topological Mott phases displaying the quantum Hall
and the quantum spin Hall effects are found for spinless and
spinful fermion models, respectively.  We present the mean-field
phase diagram and consider the effects of fluctuations within the
random phase approximation (RPA). Functional renormalization group
analysis also show that these states can be favored over the
topologically trivial Mott insulating states. 
\end{abstract}

\pacs{71.10.-w, 71.10.Fd, 71.27.+a, 71.30.+h, 73.22.Gk, 73.43.-f}

\maketitle \emph{Introduction - } Partly motivated by the
discovery of the high $T_c$ superconductivity, Mott insulators
have attracted great attention in recent years. Defined in a
general sense, interactions drive a quantum phase transition from
a metallic ground state to an insulating ground state in these systems. Most Mott
insulators found in nature also have conventional order
parameters, describing, for example, the charge-density-wave (CDW)
or the spin-density-wave (SDW) orders. However, Mott insulators
with exotic ground states, such as the current carrying ground
states have also been proposed theoretically
\cite{Affleck1988,Wen1996,Varma1999,Chakravarty2001}.
In parallel with the study of strongly correlated systems, there
has recently been a growing interest in realizing topologically
non-trivial states of matter in band insulators. In the quantum
anomalous Hall (QAH) insulators\cite{Haldane1988,Qi2006}, the
ground state breaks time reversal symmetry but does not break the
lattice translational symmetry. The ground state has a bulk
insulating gap, but has chiral edge states. In the quantum spin
Hall (QSH) insulators\cite{kane2005,bernevig2006a,bernevig2006d},
the ground state does not break time reversal symmetry, has a
bulk insulating gap, but has helical edge states, where electrons
with the opposite spins counter-propagate. The QSH state has
recently been observed experimentally in $HgTe$ quantum
wells\cite{bernevig2006d,koenig2007}.

Given the tremendous interest in finding Mott insulators with
exotic ground states, and the recent discovery of the
topologically non-trivial band insulators, it is natural to ask
whether one can find examples of topological Mott insulators,
which we define as states with bulk insulating gaps driven by the
interaction, and inside which lie topologically protected edge
states. Furthermore, electronic states in the Mott insulator
phases are characterized by topological invariants, namely, the
$U(1)$ Chern number\cite{Thouless1982} in the case of the QAH
state, and the $Z_2$ invariant\cite{kane2005A} in the case of the
QSH state.  In this letter, we report on the first example of such a 
case by systematically studying Hubbard models with repulsive
interactions on a two dimensional honeycomb lattice.  In
particular, we demonstrate how, due to the frustrated nature of
second-neighbor repulsion on this lattice, topological Mott phases
displaying the QAH and the QSH effects are generated dynamically.
We present the mean-field phase diagram, proceed to consider
effects beyond mean-field theory via a functional renormalization
group (fRG) treatment, and consider the effects arising from
fluctuations.  The two dimensional graphene sheet, which was
recently realized experimentally  \cite{Novoselov2005,Zhang2005}, 
contains the basic degrees of freedom of
our model. However, presently we do not know how to tune the
interaction experimentally so that our proposed state can be
realized.

\emph{Spinless Fermions and the QAH state - }
\begin{figure}
\includegraphics[width=3.0in]{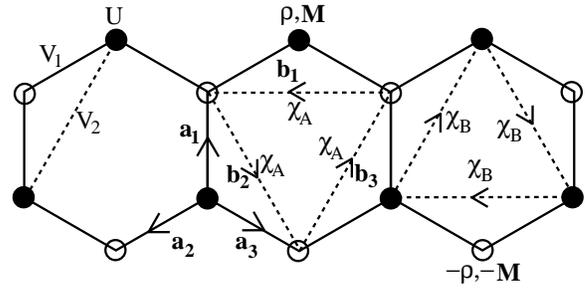}
\caption{Interactions considered in our model Hamiltonian
(left-most hexagon), Eq. \ref{hamiltonian}. Various order
parameters are shown for the A-sublattice (open circles) in the
middle hexagon and for the B-sublattice (filled circles) in the
right-most hexagon.  The CDW order parameter $\rho$ is a real
scalar, SDW order parameter $\bm M$ a real vector and the QAH/QSH
order parameters $\chi_A, \chi_B$ are complex 4-vectors.  In the
case of spinless fermions, $\chi_A, \chi_B$ are complex scalars.  $\chi_A,\chi_B$ are both defined
on the directed second neighbor links defined by $\bm b_i$.    }
\label{geometry}
\end{figure}
The model Hamiltonian for spinless fermions with nearest-neighbor
and next-nearest neighbor interactions is  written
as\begin{eqnarray} \label{hamiltonian} H=-\sum_{\langle ij \rangle}t
\left(c_i^\dagger c_j+h.c. \right)+V_1 \sum_{\langle i,j \rangle }
(n_{i}-1)( n_{j}-1) \nonumber  \\ + V_2\sum_{\langle \langle i,j
\rangle \rangle}  (n_{i}-1) (n_{j}-1) -\mu\left( \sum_in_i-N \right)
\end{eqnarray}
where $V_1 $ and $V_2$ are nearest-neighbor and
next-nearest-neighbor interaction strengths, respectively.  Since
the honeycomb lattice is bipartite, consisting of two triangular
sublattices (referred to here as A and B), nearest-neighbor
repulsion will favor a charge density wave (CDW) phase with an 
order parameter $\rho = \frac{1}{2} \left(\langle c^{\dagger}_{i
A} c_{i  A} \rangle - \langle c^{\dagger}_{i  B} c_{i  B} \rangle
\right)$ that is consistent with overall charge conservation and
describes a phase with a broken discrete (inversion) symmetry.
However, since the second neighbor interactions within a
sublattice are frustrated, CDW order will be suppressed; instead
we consider the possibility of orbital ordering by defining the
following order parameter for $i,j$ next nearest neighbors:
$\chi_{ij}=\chi_{ji}^* = \langle{c_i^\dagger c_j} \rangle$.  Let
$\bm a_1, \bm a_2, \bm a_3$ be the nearest-neighbor displacements
from a B-site to an A-site such that $\bm z \cdot \bm a_1 \times
\bm a_2$ is positive.   We also define the displacements $\bm b_1
= \bm a_2 - \bm a_3$, $ \bm b_2 = \bm a_3 - \bm a_1$, etc, which
connect two neighboring sites on the same sublattice (Fig.
\ref{geometry}).  A translational and rotational invariant ansatz
of $\chi_{ij}$ is chosen as
\begin{eqnarray} \chi_{i,i+{\bf
b}_s}=\left\{\begin{array}{c c}\chi_A = \vert \chi \vert
e^{i\phi_A},&i\in A\\ \chi_B = \vert \chi \vert e^{i\phi_B},&i\in
B\end{array}\right.
\end{eqnarray}
which are \emph{complex} scalars that live along the directed second
neighbor links.

Owing to translational
symmetry, the mean-field free energy at $T=0$ is readily obtained: 
\begin{eqnarray}
F \left( \rho, \chi, \bar{\phi}, \phi \right) &=& -  \sum_{\bm k}
\sqrt{  \vert  t(\bm k) \vert ^2 + \left( V_1 \rho +2 V_2 \vert \chi
\vert S_{\bm k  + \bar{\phi} }   S_{\phi} \right)^2 } \nonumber
\\
&&+ 3L^2 \left( V_1  \rho^2 + 2V_2 \vert \chi \vert ^2 \right)
\end{eqnarray}
We have defined the following quantities: $t(\bm k) = \sum_{n=1}^3
\exp \left( i \bm k \cdot a_n \right)$, $ \bar{\phi} = \left( \phi_A
+ \phi_B \right)/2, \phi = \left(\phi_A - \phi_B \right)/2,S_{\bm k+
\bar{\phi}}=\sum_{n=1}^3 \sin {\left(\bm k\cdot b_n + \bar{\phi}
\right)}, S_{\phi} = \sin{\phi} $.
Thus, the next-neighbor hopping
amplitudes are purely real only when both $\phi = 0$ and $\bar{\phi}
= 0$.

When both $\rho$ and $\chi= 0$, and at half-filling, the system
remains a semi-metal consisting of two Fermi points $\bm K_{\pm}$
which obey $\bm K_{\pm} \cdot \bm b_i = \pm 2 \pi/3$ and the density
of states vanishes linearly; the dispersion in the vicinity of these
so called Dirac points is governed by a 2D massless Dirac
Hamiltonian in $\bm k$-space . The CDW phase corresponds to an
ordinary insulator with a gap at the Fermi energy. As for the order
parameter $\chi$, which describes the second-neighbor hopping, its
\emph{phase} relative to the nearest neighbor hopping amplitude
plays an important role in determining its properties: while a
non-zero $Re(\chi)$ merely shifts the energy of the Dirac points, a
non-zero imaginary part $Im(\chi)$ \emph{opens a gap} at the Fermi
points.  Thus, when the system remains at half-filling, it is more
favorable to develop  purely imaginary next-neighbor hopping
amplitudes; such a configuration corresponds to a phase with
\emph{spontaneously broken time-reversal symmetry}.

\begin{figure}
\includegraphics[width=2.5in]{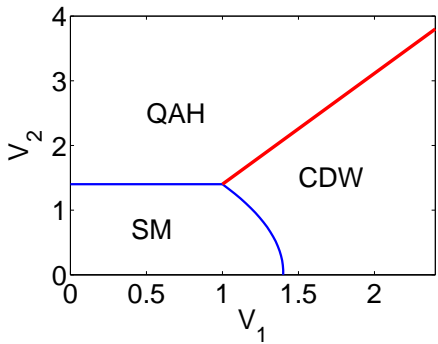}
\caption{Phase diagram for spinless fermions ($t=1$).  The
semi-metallic (SM) state that occurs at weak-coupling is separated
from the CDW and the topological QAH states via a continuous
transition (blue curve). The line separating the QAH and CDW marks a
first-order transition (shown in red), which terminates at a bi-critical point.}
\label{pd2}
\end{figure}

To see whether such a phase can be favored, we minimize the
free-energy and arrive at the following self-consitent
equations:\begin{equation} \rho = \frac{1}{2 L^2} \sum_{\bm k}
\frac{V_1 \rho + 2V_2\chi S_{\bm k + \bar \phi} S_{\phi}}{\sqrt{
\vert t(\bm k) \vert^2
 +\left(V_1 \rho + 2V_2\chi S_{\bm k + \bar{\phi} } S_{\phi}\right)^2 } }
\end{equation}
\begin{equation}
\chi = \frac{S_{\phi}}{6L^2} \sum_{\bm k} \frac{S_{\bm k +
\bar{\phi} }   \left(V_1 \rho + 2V_2 \chi S_{\bm k + \bar \phi }
S_{\phi} \right)}{\sqrt{ \vert t(\bm k) \vert^2 +\left(V_1 \rho +
2V_2 \chi S_{\bm k  + \bar{\phi} } S_{\phi}\right)^2 }
}
\end{equation} When $\chi = 0$, it is easy to see from the first
equation above that CDW order develops continuously at a critical
value $V_{1c}$ given by
\begin{equation}
\frac{1}{V_{1c}} = \frac{1}{2L^2} \sum_{\bm k} \frac{1}{\vert t(\bm k) \vert},
\end{equation}
Due to the vanishing density of states (DOS) near the Fermi points,
there is no instability towards CDW formation with infinitesimal
interactions.  Interestingly, the self-consitent equation for
$\chi$ shows that a non-trivial self-consistent solution can only
occur when $\phi \ne 0$; a detailed investigation of these
equations \cite{Raghu2007} show that when $V_1 = 0$, beyond a
critical value of $V_{2c}>0$, which satisfies\begin{equation}
\label{gapeqQAHE} \frac{1}{V_{2c}} = \frac{1}{3L^2} \sum_{\bm
k}\frac{S^2_{\bm k + \bar{\phi}} }{\vert t_{\bm k} \vert},
\end{equation}
a phase in which $\vert \chi \vert > 0$ ,$\bar{\phi} = 0$, and
$\phi = \pm \pi/2$ is favored.  Such a phase is also stable at
finite $V_1$ and is thus does not require fine-tuning (see Fig.
\ref{pd2}). In this phase, the system acquires purely imaginary
second-neighbor hoppings with a chirality which is determined by
the sign of $\phi$. There is a discrete symmetry breaking
corresponding to choosing $\phi = \pm \pi/2$, each of which breaks
time-reversal symmetry. The band insulator version of the CDW
state was considered in Ref. \cite{Semenoff1984}, while the QH
state on a honeycomb lattice was considered in Ref.
\cite{Haldane1988}. As discussed in Ref. \cite{Haldane1988} the
QAH phase is a topological phase in which the filled states form a
band which has a non-zero topological Chern number
\cite{Thouless1982} and is an integer quantum Hall effect phase
that is realized \emph{without} Landau levels. Similar states can
be constructed from magnetic semiconductors\cite{Qi2006}. We shall
generally refer to QH states without Landau levels, and with full
lattice translation symmetry as the \emph{Quantum Anomalous Hall}
(QAH) states. In our case, the topologically non-trivial gap for
the QAH state arises from many-body interactions rather than
single particle physics, and we shall refer to such states as
topological Mott insulators.

We have obtained the complete  phase diagram in the $V_1-V_2$
plane; within mean-field theory, there is a continuous transition
from the semi-metal to either the CDW or the QAH phase and there
is also a first-order transition from the CDW to the QAH phase. By
integrating out the fermionic fields, it is possible to construct
a Landau-Ginzburg (LG) theory expansion near the nodal region
where the order parameters are vanishingly small.  Due to the
linear dispersion of the Fermi points, the LG free-energy contains
anomalous terms of the form  $\vert \rho \vert^3$ and $\vert
Im(\chi) \vert^3$ arising from linear dispersion in the vicinity
of the Fermi points \cite{Raghu2007}.  The significance of such
terms is that even within mean-field theory, the CDW order
parameter, for instance, grows as $(V_1 - V_{1c})$ rather than the
usual $\left( V_1 - V_{1c} \right)^{1/2}$ \cite{Sorella1992}.
Furthermore, the Landau-Ginzburg theory describing the competition
between the CDW and QAH phases confirms the existence of the first
order line between these two phases that terminates in the
bicritical point leading into the semimetal phase
\cite{Raghu2007}.

\begin{figure}
\includegraphics[width=3.0in]{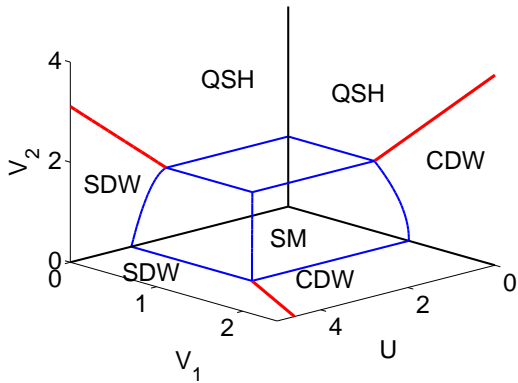}
\caption{Complete mean-field phase diagram for the spinful model.  The transitions from the semimetal (SM) to the insulating phases are continuous, whereas transitions between any two insulating phases (red lines) are first-order.  
} \label{pd3}
\end{figure}

\emph{Spinful fermions and the QSH state -} Next, we take into
account the spin degrees of freedom and include an onsite Hubbard
repulstion in our model Hamiltonian ($\mu = 0$):
\begin{eqnarray}
H=-\sum_{\langle ij \rangle  \sigma}t\left(c_{i \sigma}^\dagger c_{j
\sigma}+h.c. \right)+U\sum_i n_{i \uparrow} n_{i \downarrow}
\nonumber \\ + V_1\sum_{\langle i,j \rangle }   (n_{i}-1)( n_{j}-1)
+ V_2\sum_{\langle \langle i,j \rangle \rangle}  (n_{i}-1)(
n_{j}-1)\label{hamilton}
\end{eqnarray}
where $n_i = n_{i \uparrow} + n_{i \downarrow}$.  Since the
Honeycomb lattice is bipartite, onsite repulsion gives rise to a
spin density wave phase (SDW); a standard decomposition of the Hubbard term 
introduces the order parameter $\bm M$ describing an
antiferromagnetic SDW: $\bm M = \frac{1}{2} \left( \langle \bm
S_{iA} \rangle-\langle \bm S_{iB} \rangle \right)$ .  As in the
spinless case, nearest-neighbor repulsion favors a CDW.  However,
there are several possible phases due to second-neighbor
repulsion.   Again, since the second-neighbor repulsion is
frustrated, we are again led to the possibility of a topological
phase similar to the QAH.  However, the spin degrees of freedom
introduce two possibilities (translation invariance along with
spin conservation eliminate other possibilities): 1) \emph{two
copies} of QAH states - i.e. the chirality of the
second-neighbor hopping is the \emph{same} for each spin projection, 2)
the QSH state, where the chiralities are \emph{opposite} for each
spin projection.   The latter possibility breaks a
\emph{continuous} global $SU(2)$ symmetry associated with choosing
the spin projection axis;  however, \emph{time-reversal symmetry
is preserved}. The QSH state on the honeycomb lattice was
considered in Ref. \cite{kane2005}, where the insulating gap
arises from the microscopic spin-orbit coupling. It was shown
latter that the magnitude of of the spin-orbit gap is negligibly small
in graphene \cite{Min2006,Yao2007}. In our case, the insulating gap is generated
dynamically from the many-body interaction. In this sense, our
effect can be viewed as an example of dynamic generation of
spin-orbit interaction\cite{Wu2004}.
  Introducing the Hubbard-Strataovich fields (sum over repeated indices implied)
$\chi^{\mu}_{ij} = c^{\dagger}_{i \alpha} \sigma^{\mu}_{\alpha
\beta} c_{j \beta}$, $\mu = 0 \ldots 3$, where $\sigma^{\mu} =
\left( 1, \bm \sigma \right)$, the next-neighbor interactions can
be recast using the identity $ (n_{i } -1)(n_{j}-1) =1  -
\frac{1}{2} \left( \chi^{\mu }_{ij} \right)^{\dagger}
\chi^{\mu}_{ij} $. Physically, if $\langle \chi^0 \rangle \neq 0$,
then we are in the QAH phase.  If, on the other hand, one of the
vector components $\langle \chi^{i} \rangle \neq 0$, then we are
in the QSH phase.  A translationally invariant decomposition of
the next-neighbor interactions via $ \langle \chi^{\mu}_{i,i+\bm
b_s} \rangle = \chi^{\mu}e^{i \phi^{\mu}_A} , i \in A $ (and
similarly for the other sublattice) gives rise to 
a $4 \times 4$ Hamiltonian is readily diagonalized in
a tensor product basis $\bm \sigma \otimes \bm \tau$, where $\bm
\sigma$ and $\bm \tau$ are Pauli matrices in spin and sublattice
space, respectively. This way, each phase corresponds to a
particular non-zero expectation value of a fermion bilinear
$\sum_{\vec{k}}\Psi_{\vec{k}}^\dagger \hat{d}(\vec{k})
\Psi_{\vec{k}}$, where $\hat{d}(\vec{k}) \propto \tau^3$ for the
CDW and QAH, $\hat{d}(\vec{k}) \propto \sigma^3\tau^3$ for SDW and
QSH.  A detailed study of the free-energy at $T=0$ and its saddle
point solutions \cite{Raghu2007} produces the phase diagram shown
in Fig. \ref{pd3}.  In addition to the ordinary CDW and SDW
insulating phases, there is a phase for $V_2 > V_{2c} \approx 1.2
t$ in which the 4-vector is purely imaginary (as in the spinless
case), collinear, and staggered from one sublattice to the next:
$\langle \chi^{\mu}_{i i+\bm b_n, A} \rangle = - \langle
\chi^{\mu}_{i i+\bm b_n, B} \rangle$, and both QAH and QSH are
equally favorable ground states, having identical free energies
within mean-field theory.  Additionally, there is never a
coexistence of both QAH and QSH phases; indeed, a Landau-Ginzburg
treatment in this region explicitly shows the absence $SO(4)$
symmetry of the vector $\chi^{\mu}$.  This occurs due to the
difference of the manner in which $\chi^0$ and $\vec{\chi}$ are
coupled to the fermionic fields - which favors either a phase with
broken $Z_2$ symmetry (QAH) or with broken $SU(2)$ symmetry, but
never both simultaneously \cite{Raghu2007}.

Quantum fluctuations, however, lift the mean-field degeneracy
between the QAH and QSH phases.  To quadratic order in quantum
fluctuations (RPA) about the QSH phase , we obtain an effective
action $ S_{eff} = \sum_{\vec{k}} \delta \chi^{\mu}(\vec{k},
\Omega) K_{\mu \nu} (\vec{k}, \Omega ) \delta \chi^{\nu}
(-\vec{k}, -\Omega)$ which shows the presence of six modes (2
longitudinal and 4 transverse modes), and 2 of the transverse
modes correspond to degenerate Goldstone modes whose velocity is
proportionality to the Fermi velocity $v \approx v_f = 3t/2\vert
\bm a \vert $.  Thus, the zero-point motion associated with these
gapless modes, lowers the free energy of the QSH state relative to
the QAH state.

\begin{figure}
\includegraphics[width=3.5in]{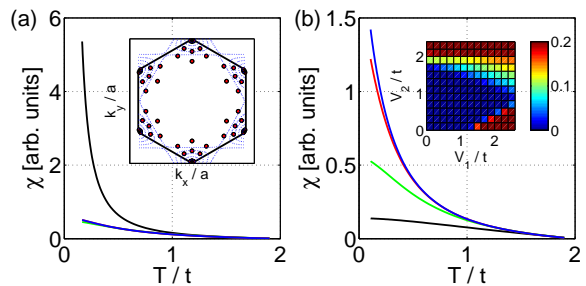}
\caption{{\em a)} Data for $U$=0, $V_1$=1.4$t$, $V_2$=0.  Susceptibilities of each phase vs. $T$ are shown: CDW (black);  SDW (green); QAH(red) and QSH (blue).{\em b)} Same for $U$=0, $V_1$=0, $V_2$=1.8$t$ (QSH instability).  The QSH phase has a larger susceptibility than QAH.   {\it Inset}: fRG phase diagram at $U$=0, indicating SM (blue) and insulating (red) regions (CDW dominates at large $V_1$, QSH at large $V_2$.)   The colorbar correspond to $T_c$ below which the insulating phases develop in fRG.}
\label{rgfig}
\end{figure}

\emph{Renormalization Group Analysis - } Mean field
theory generally starts with a given, in a sense, biased Ansatz,
and investigate the self-consistency of the mean field solution.
Therefore, it is important to investigate the topological Mott
states with a method without any a priori bias. Next we go beyond
mean-field theory and RPA using the temperature($T$)-flow
functional renormalization group
(fRG)\cite{Honerkamp2001}\cite{Honerkamp2001A}.  
In this scheme, we discretize the $\vec{k}$-dependence of the
interaction \cite{Zanchi2000}
and consider all possible scattering processes between a set of initial
and final momenta that occur between
points on rings around the Dirac points (inset of Fig. \ref{rgfig}).
Starting with $T_0 \sim 2t$, the
temperature $T$ is lowered, and a flowing (renormalized) interaction
$V_T$ is obtained by the
coupled summation of the $T$-derivatives of {\em all} one-loop channels.
Due to this, the method is
unbiased and goes beyond the mean-field-level.  
Applying the scheme to the Hamiltonian, Eq.
\ref{hamilton}, we search for {\em
  flows to strong coupling}, where for a low temperature $T_c$ certain
components of $V_T $ become large. Then the approximations break
down, and the flow is stopped. Information on the low-$T$ state is
obtained from analyzing which coupling functions grow most
strongly and from susceptibilities for static external fields
coupling to the various order parameters.  In this scheme, a
tendency towards ordering at a finite vector $\bm Q$ can be
detected as a growth of the associated vertex $V_T$.  However, we
have found that largest couplings occur at $\bm Q = 0$, which
strongly supports the mean-field results presented above.

For onsite and nearest-neighbor repulsions $U>U_c \approx 3.8t$
and $V_1> V_{1c} \approx 1.2t$, the flow to strong coupling is
either an SDW instability for dominant $U$ or a CDW instability
for dominant $V_1$, in good agreement with a
1/N-study\cite{Herbut2006} and
Quantum-Monte-Carlo\cite{Sorella1992}. For more details, see Ref.
\cite{Honerkamp2007}. If we include a sufficiently strong
second-nearest-neighbor repulsion $V_2>1.6t$, the flows change
qualitatively; there is a leading growth of the QSH
susceptibility. In Fig. \ref{rgfig} a) and b) we compare the
$T$-flows of various susceptibilities for $V_1>V_2$ and for
$V_2>V_1$. For the latter case, the QSH susceptibility grows most
strongly toward low $T$, followed by the QAH susceptibility, which
is consistent with the RPA treatment of the Goldstone modes in the
QSH. The QSH phase remains stable even when a moderate onsite
interaction of $U=t$ or $U=2t$ is introduced. Hence the global
structure of the mean-field phase diagram is confirmed by the fRG
results. Note however that the slope of the lines of critical
$V_1$ versus $V_2$ differs. We interpret this a competition effect
captured by the fRG, where $V_2$ decreases the CDW tendencies
induced by $V_1$.

\emph{Discussion - } We have shown that topological phases
displaying the QAH and QSH effects can be generated from strong
interactions - thus, we refer to these phases as \emph{topological
Mott insulators}. Both phases have associated with them
conventional order parameters which develop continuously at the
quantum critical phase transition from the semi-metallic state.
However, these states are also described by topological quantum
numbers which jump discontinuously at the transition. Although
the interaction strengths needed to produce these phases are
strong, we expect that our mean-field, RPA and fRG treatment to
provide strong evidence for the existence of a topological Mott
insulator: the Stoner criterion, which states that perturbation theory 
breaks down when interaction strengths are comparable to the 
inverse of the effective DOS, suggests that perturbation theory 
is more robust in our system due to the vanishing of the 
DOS at the Fermi points.  
An open issue remains to find other possible realizations
of the phases described here. Furthermore, the nature of the low energy
effective coupling of the gapless bulk Goldstone modes with the gapless
\emph{edge} degrees of freedom, and the possibility of
fractionalized excitations in these phases are interesting
open issues that we leave for future work.

\section{Acknowledgments}
We are grateful to D. P. Arovas and S. A. Kivelson for insightful discussions.   
This work is supported by the NSF under grant numbers DMR-0342832, 
the US Department of Energy, Office of Basic Energy Sciences
under contract DE-AC03-76SF00515, BaCaTec (C.H.),
and the Stanford Institute for Theoretical Physics (S.R.).

\bibliography{tmi}

\end{document}